# Evaluation of the Antibacterial and Wound Healing Properties of a Burn Ointment Containing Curcumin, Honey, and Potassium Aluminium

Mahsa Shahbandeh[1,2], Mahsa Amin Salehi[3], Maryam Soltanyzadeh[4], Mehrnaz Mirzaei[5], Ali Maleki[6], Abdolkarim Chehregani rad[3], Mohammad Javad Fatemi[2], Reza Mirnejad[7], Mostafa Dahmardehei[2]

[1] Terita Sana Darman Company, Tehran, Iran.
[2] Burn Research Center, Iran University of Medical Sciences, Tehran, Iran
[3] Department of Plant Science, Faculty of Sciences, Bu-Ali Sina University, Hamedan, Iran
[4] Department of Cell and Molecular Biology, Faculty of Biological Sciences, Kharazmi University, Tehran, Iran.
[5] Department of Microbiology, Tehran Medical Sciences Branch, Islamic Azad University, Tehran, Iran.
[6] Physics department, University of Ottawa, 25 Templeton St., Ottawa, ON, Canada
[7] Molecular Biology Research Center, Systems Biology and Poisoning Institute, Baqiyatallah University of Medical Sciences, Tehran, Iran.

**Abstract**

Burn wounds can severely trouble the health system and life quality of patients. The present study aimed to analyze the synergistic healing properties of curcumin, honey, and potassium alum substances merged in a newly-devised burn ointment on second-degree burn wounds in rats. The MIC and MBC tests on 200 clinical isolates of *Pseudomonas aeruginous* are compared to imipenem *in vitro*. Their killing time and cytotoxicity are also studied using a standard isolate of *P. aeruginous*, fibroblast stem cells (FSC) and mouse embryonic fibroblasts (MEF). Furthermore, histopathological and histomorphological assessments are conducted on 150 male Wistar rats whitin four experimental groups to evaluate the efficiency of the prepared burn ointment. We found a significant wound healing in both macroscopical observations and microscopical evaluations. Both curcumin and honey show strong antimicrobial effects with no cytotoxicity. Also, the histopathological results present a considerable and comparable wound re-epithelization in the a group of rats treated with both honey and curcumin after 7 days. The burn ointment containing curcumin, honey, and potassium alum show considerable efficacy in accelerating the healing of experimentally-induced burn wounds in animals. Th novel onement product is propose as a powerful alternative for the topical treatment of burn injuries.

## 1 Introduction

Burn wounds can be difficult-healing traumas causing much anticipatory anxiety and emotional distress due to their painful injury and therapeutic procedures[1–3]. Either chemical agents, heat, cold, electricity, radiation, or friction can cause burn wounds of four degrees (I to IV)[4]. They are at high risk of infection, scar-leaving, and skin contractures due to myofibroblast accumulation, excessive collagen fiber production, and low vascular blood supply during the healing process[1,5]. The oxidative stress in the wound area is followed by a general healing process including inflammation, granulation acting as a stratum for cell migration and recruitment (e.g., fibroblasts, endothelial cells, and macrophages), matrix deposition, re-epithelialization, and vascularization as a necessity for tissue remodeling[6,7]. Re-epithelialization occurs through migration, proliferation, and differentiation of epithelial cells[8,9].

All burn wounds need precise observations due to their slow healing process and risk of infection[10]. Also, exposing parts of them are highly at risk of bacterial infections especially with *Pseudomonas aeruginosa* that can disrupt the healing processes and develop life-threatening complications[11]. The extensive drug-resistant (XDR) or multi-drug resistant (MDR) strains of *P. aeruginosa* are associated with a high mortality rate. Carbapenems such as imipenem are the current standard antibiotics used against XDR *P. aeruginosa*[12]. However, carbapenem-resistant Pseudomonal infections are rapidly growing. Therefore, several studies are focused on finding effective compounds against MDR and XDR *P. aeruginosa* strains. Honey and curcumin are two candidates studied for antibacterial and wound healing activities unadulterated or in combination with other materials [13,14]

Several topical medicines are being developed to promote the healing process and subside the scar [15]. Probiotic therapy using *Lactobacillus acidophilus* in an ointment formulation [16] phytomedicine [17], and traditional therapies revived in the form of modern products including herbal- (e.g., curcumin) and animal- (e.g., honey and propolis) originated materials are some recently studied solutions for the burn wounds. They are expected to provide more accessible and purchasable alternatives with minimized adverse effects[18]. The medicinal extract of *Delonix regia* [19], *Arnebia Euchroma*, *Lawsonia inermis* [20]*Aloe vera* [21] Capsicum [22], marigold (*Calendula officinalis*)[23], and *Avena* spp.[24] are some herbal-originated products studied for their burn wound healing, anti-inflammatory, and analgesic activities. The irritating pain or systemic complications associated with some above-mentioned derivatives have motivated us to develop a therapeutic product able to simultaneously relieve the pain and hyper-inflammatory conditions and accelerate the healing process using honey, curcumin and Aluminum potassium phosphate [25].

Honey is a viscous solution with high healing potential that has been traditionally used and widely studied as a natural care product for various types of wounds due to the antimicrobial, antioxidant, anti-inflammatory, immunomodulatory, and analgesic properties[20,26]. Curcumin is a polyphenolic phytochemical extracted from turmeric and widely studied for its wound healing, analgesic, antioxidant, and anti-inflammatory properties [18,27]. Curcumin owes its healing properties mainly to the three components of diferuloylmethane, bisdemethoxycurcumin, and demethoxycurcumin [28]. Abbas *et al*. [11] showed the synergistic antibacterial effect of curcumin in cross-link with chitosan-combined polyvinyl alcohol (PVA) membrane. El-Refaie *et al*. [13] used curcumin in a nanogel-core hyaluronic acid liposomal structure (hyaluosomes) and could modulate its hydrophobicity, increase its skin absorbance, prevent its degradation, and increase its skin penetration. It has been revealed that the mixture of curcumin and *Hypericum perforatum* can be a promising alternative to silver sulfadiazine [29].

Aluminum potassium phosphate (also called potassium alum, potash alum) with the molecular formulation $KAl(SO_4).12H_2O$ is a neutral crystalline mineral usually used as a food preservative agent. Considering its strong antibacterial properties[30], Alasbahi et al. have shown selective healing effectiveness for potassium alum specifically on wounds suggested through contracting the wound by its adstringent property [31]. This greek ancient remedy has also shown other medicinal potentials such as antibacterial properties and hemostaticity causing its wide application for several applications including wound healing [32].

This study aimed to develop a burn ointment that provides both healing and relieving properties for burn-wound patients. For this purpose, the antibacterial and wound healing properties of our product containing curcumin, honey, and potassium alum are evaluated both in vitro and in experimentally-prepared animal models. Where the MIC and MBC of the proposed mixture on *Pseudomonas aeruginous* are compared to the imipenem in vitro. Also, 150 male rats are exposed to the current natural product, ointment base, and zinc oxide ointment. The burn treatment efficiency of each material is studied by histopathological and histomorphological assessments. To the knowledge of the authors, this is the first time that honey, curcumin, and potassium alum are used in a single burn ointment product.

## 2 Material and Methods

### 2.1 Materials

All chemicals and reagents including curcumin and bacterial and cellular culture media are prepared from Sigma-Aldrich (USA). FSC and MEF were purchased from Pasteur Institute of Iran (IPI) and Honey from PONIK pharmaceutical co., Iran. Both honey and curcumin used in this study were of the pharmaceutical grade and purification analyses were prepared by the suppliers.

### 2.2 Bacterial Strains and Culture

Antibacterial properties of curcumin, honey, and potassium alum were examined against the standard strain *Pseudomonas aeruginosa* ATCC 27853 and 200 clinical isolates of *P. aeruginosa*. *P. aeruginosa* ATCC 27853 was purchased from IPI. The clinical isolates were collected from urine, blood, endotracheal tube, sputum, trachea, and wounds in the Central Laboratory of Iran University of Medical Sciences as was explained in our previous study [12]. All bacteria strains were grown on three culture mediums (blood agar, Trypticase soy broth, and MacConkey agar) at 37°C for 24 hours. Then, bacterial suspension was prepared in BPS and adjusted to 0.5 McFarland to be used as inoculum [33].

### 2.3 Antibiogram Test

For antibiotic resistance analysis, 200 clinical isolates were tested with 21 different antibiotics using the disc diffusion method based on the Clinical & Laboratory Standards Institute (CLSI), "Performance Standards for Antimicrobial Disk Susceptibility Tests", 13th Edition [34]. Isolates were considered as sensitive if they showed no or partial growth. Otherwise, they were considered as the resistant [33].

### 2.4 MIC and MBC Evaluations

MIC and MBC of both active materials (curcumin, honey, and potassium alum) and imipenem (as control) were evaluated for the clinical isolates of *p. aeruginosa* using the standard guidelines of the CLSI. Briefly, an initial bacterial inoculum containing $1.5 \times 10^8$ CFU/mL (0.5 McFarland) of isolates was inoculated in cation-adjusted Mueller-Hinton broth (CAMHB; BD Diagnostic Systems, adjusted to pH = 5.9) in a 96-well microplate. The inoculated wells were exposed to serial micro dilutions of curcumin, honey, potassium alum, and imipenem. The endpoints were determined when no turbidity was observed in the well. Then, the microplate was incubated for 24 h at 37°C. After 24 h of incubation, the absorbance of each well was measured at 600 nm using a microtiter plate reader. The lowest concentrations that inhibited the growth of 50% and 99% of bacteria were considered as MIC and MBC, respectively. Each experiment was repeated three times[35].

### 2.5 Time Killing Assay

For analyzing the post-treatment bacterial viability and determining the minimum time needed for reaching an inhibitory or bactericidal effect, viable *P. aeruginosa* bacteria were counted at various concentrations of curcumin, honey, and potassium alum. The killing time curve was delineated for the survived *P. aeruginosa* isolates post-challenged by curcumin, honey, potassium alum, and imipenem at their MIC values (8 mg/µL, 16 mg/µL, and 64 mg/µL, respectively). For this purpose, the standard isolate of *P. aeruginosa* at the exponential phase of growth was obtained by culturing in Muller–Hinton broth (MHB). A standard 0.5 McFarland suspension ($1.5 \times 10^8$ CFU/mL) of each was aliquoted in a microplate and diluted to $5 \times 10^5$ CFU/mL (1/300). The diluted bacteria suspension is mixed with curcumin, honey, potassium alum, and imipenem and incubated at 37°C. Each aliquot was sampled at 0, 0.5, 1, 2, 4, 8, 16, 18, and 24 h, and cultured on Muller-Hinton agar (MHA) for another 24 h at 37°C. Colony count was exerted after 24 h to obtain the CFU of each aliquot at either 0, 0.5, 1, 2, 4, 8, 16, 18, or 24 h[19].

**2.6 Cell Culture**

The toxicity of the prepared burn ointment and its main ingredients (curcumin, honey, and potassium alum) on the eukaryotic cells was evaluated using FSC and MEF. Cells were cultured in 48-well microplates in the Dulbecco's Modified Eagle Medium (DMEM) supplemented with 1% fetal bovine serum (FBS) and incubated for 24 h at 37°C and 5% $CO_2$[21]. Then, the cells were used for further experiments.

**2.7 MTT Test**

FSCs and MEFs were treated with several concentrations of curcumin, honey, and potassium alum, and their proliferation ability was measured via assessing the cell metabolic activity by 3-(4,5-dimethylthiazol-2-yl)-2,5-diphenyltetrazolium bromide (MTT) assay. First, cells were dispensed in a 48-well culture plate ($1 \times 10^4$ cells/well) and incubated at 37°C and 5% $CO_2$. 24 hours later, different concentrations of curcumin, honey, and potassium alum (25, 51, and 102 µg/mL) were transferred to each well, and the vitality of cells was evaluated by MTT assay after 1, 3, and 5 days post-treatment. In this regard, after 24, 72, and 120 hours, the culture suspension was replaced by the MTT solution (5 mg/mL) and incubated at 37°C for 3 hr. At each specified time, the supernatant was removed and the remained formazan crystals were dissolved in dimethyl sulfoxide (DMSO). Finally, the absorbance of each well was measured by a microplate reader (BioTek Instruments) at 540–630 nm[12].

**2.8 Morphological Surveys**

To better understand the morphological changes and probable apoptotic effects of the ointment on eucaryotic cells, the growth of FSC and MEF in presence of different concentrations (25, 51, and 102 µg/mL) of ointment were observed under acridine orange/ethidium bromide (AO/EB) double staining after 1, 3, and 5 days. For staining, fluorescent dyes including ethidium bromide acridine orange (AO, 100 µg/mL) and (EB, 100 µg/mL) were added to the wells and the living and dead cells were investigated with a fluorescence microscope (Nikon, Japan). The cell injury status was evaluated using morphological criteria. Green nuclear staining was considered as cells containing normal nuclear chromatin, and yellow to red nuclear staining was considered as cells with fragmented nuclear chromatin due to apoptosis [36]

**2.10 Animals**

150 adult male Wistar rats (280 and 300 gr, 8 – 10 weeks) were prepared from the Animal Laboratory Service of Pasteur Institute of Iran, Tehran, Iran. Rats were acclimated for one week before the experiment initiation. All animal experiments were conducted under aseptic conditions according to the guidelines of the Care and Use of Laboratory Animals [37]. Rats were kept in separate standard Plexiglas cages with access to water and food at libitum (standard pelleted chow). The environmental conditions of housing included a temperature of 23 ± 3°C, 50 ± 5% humidity, and a 12/12 h dark/light cycle. We hereby confirm that the study is reported in accordance with ARRIVE guidelines (https://arriveguidelines.org).

**2.11 *In vivo* Study**

For in vivo evaluation of antibacterial properties of the present burn ointment, 150 adult male rats were experimentally induced for burn wound infection. The present study design was approved by the Animal Ethics Committee of Baqiyatallah University of Medical Sciences, Tehran, Iran[IR.Bmsu.REC.1396.880]. We declare that all methods were performed in accordance with the relevant guidelines and regulations. For this purpose, rats were first generally anesthetized using the intramuscular administration of ketamine (60 mg/kg) and xylazine (10 mg/kg). The dorsum hair of rats was shaved using an electric clipper and the area was disinfected with 70% alcohol. A deep second-degree burn was induced on the animals' dorsal skin by pressing an iron stamp with 2 × 4 cm dimensions kept in 100°C boiling water for 15 s on the animals' skin for 10 sec. Then, the pseudomonal infection was induced by subcutaneous injection of the 0.5 McFarland suspension of *Pseudomonas aeruginosa* (20 μL). Rats were randomly divided into four groups. Group 1 received a daily bandage of the prepared burn ointment (treatment group), group 2 received a daily bandage containing the base of prepared ointment (base group), group 3 received a daily bandage of zinc oxide ointment (positive control group), and finally, group 4 received no bandage during three weeks (negative control group)[13].

Burn wounds were sampled every day and the samples were cultured in the Muller-Hinton-Agar (MHA) culture medium and the colony counts were measured. Burn wounds were washed by physiologic serum and photographed using an ordinary digital camera before daily bandage for being evaluated in terms of the healing process and scar remains.

**2.12 Microscopic Evaluation**

After 7, 14, and 21 days post-treatment, four rats from each group were sacrificed using chloroform. Their skin biopsies were harvested and fixed in the 10% neutral buffered formalin (pH = 7.26) for 48 h. For preparing the histological slides, the fixed tissue specimens were dehydrated in alcohol series, cleared in xylene, embedded in paraffin, and sectioned at 5 mm thickness. All sections were stained with hematoxylin and eosin (H&E) and Masson's trichrome (MT). The histological slides were examined by an independent expert blinded to the experiment, using light microscopy with a magnification of ×40 and ×100 (Olympus BX51; Olympus, Tokyo, Japan) to assess the histopathological changes. Four parameters of epithelialization, inflammatory cell infiltration, collagenization, and granulation were surveyed and compared in the four groups [1].

**2.13 Histomorphometric Analysis**

Three parameters of collagenization, angiogenesis, and epithelialization were analyzed as follows. Collagen density was measured using the computer software Image-Pro Plus® V.6 (Media Cybernetics, Inc., Silver Spring, USA). Angiogenesis was assessed by counting the number of vessels in 5 sections at 400x magnification of the light microscope. The mean of counts in each group was reported. Epithelialization was semi-quantitatively assessed after 7, 14, 21 days, on a 5-point scale: 0 (no epithelialization), 1 (25% epithelialization), 2 (50% epithelialization), 3 (75% epithelialization), and 4 (100% epithelialization). The results of these parameters evaluation were validated with comparative analysis by an independent expert who was blinded to the treatment [38].

## 2.14 Statistical Analysis

All assays were run in triplicate and data are presented as mean ± standard deviation (SD). All quantitative data analyses were conducted using ANOVA (non-parametric) and Kruskal Wallis (parametric) analyses. Results with *P*-values < 0.05 were considered as statistically significant. Statistical analyses were performed using the SPSS software, version 20.0 (SPSS, Inc, Chicago, USA).

## 3 Results

### 3.1 Antibiogram Evaluation

All 200 clinical *P. aeruginosa* isolates (100 %) showed to be resistant to at least one antibiotic. The antibiotic susceptibility profiles of the isolates are presented in Table 1. All clinical isolates were tested with all antibiotics (as is shown in table 2). Accordingly, over 50 percent of the clinical isolates were resistant to ceftazidime, gentamicin, and imipenem. Also, azithromycin, ofloxacin, colistin, nitrofurantoin, ampicillin, norfloxacin, and cefazolin were the only antibiotics showing > 80% bacteriostatic efficiency among the isolates.

*Table 1. Antibiotic resistance of P. aeruginosa clinical isolates*

| Antibiotics | Abbreviation | Sensitivity | Resistance |
|---|---|---|---|
| Ceftazidime | CAZ | 36% | 64% |
| Gentamicin | GM | 43.5% | 56.5% |
| Imipenem | IMI | 47.5% | 52.5% |
| Ceftriaxone | CRO | 59.5% | 40.5% |
| Ciprofloxacin | CP | 59.5% | 41% |
| Ampicillin-sulbactam | SAM | 62.5% | 37.5% |
| Amikacin | AMK | 65.5% | 34.5% |
| Meropenem | MEN | 66.5% | 33.5% |
| Trimethoprim-sulfamethoxazole | SXT | 70% | 30% |
| Piperacillin-tazobactam | PITZ | 74.5% | 25.5% |
| Cefoxitin | CFO | 76.5% | 23.5% |
| Piperacillin | PIP | 76.5% | 23.5% |
| Cefepime | CPM | 80% | 20% |
| Cefotaxime | CTX | 80% | 20% |
| Ofloxacin | OFX | 82.% | 18% |
| Colistin | CO | 85% | 15% |
| Nitrofurantoin | FM | 85.5% | 14.5% |
| Ampicillin | AMP | 88.5% | 11.5% |
| Azithromycin | AZM | 89.5% | 10.5% |
| Norfloxacin | NOR | 90% | 10% |
| Cefazolin | CZ | 91.5% | 8.5% |

### 3.1 MIC/MBC Values

The results of MIC and MBC tests of the additives (curcumin, honey, and potassium alum) and imipenem (as comparative material) on *P. aeruginosa* are summarized in Table 2. It is shown that curcumin and honey represented a higher antibacterial power of one-fourth and one-eighth in comparison with imipenem, respectively. Although, the antibacterial results of the potassium alum are comparable to those of imipenem. however, the synergistic effect of the three materials is substantial. Accordingly, the MIC and MBC against *P. aeruginosa* ATCC 27853 were at least 128 and 256 µg/mL for imipenem, 16 and 32 µg/mL for curcumin, 8 and 16 µg/mL for honey, and 64 and 128 for potassium aluminum, respectively. The best antibacterial results were obtained from the combination of curcumin, honey, and potassium alum (Table 2).

*Table 2. Antibacterial analyses of curcumin, honey, potassium alum, and their combination. MIC and MBC of isolates and standard of P. aeruginosa*

| Treatments | Clinical isolates | | Standard strain | |
|---|---|---|---|---|
| | MIC (µg/mL) | MBC (µg/mL) | MIC (µg/mL) | MBC (µg/mL) |
| Imipenem | 256 | 512 | 128 | 256 |
| Curcumin | 16 | 32 | 16 | 32 |
| Honey | 8 | 16 | 8 | 16 |
| Pottasium alum | 64 | 128 | 64 | 128 |
| Synergistic honey + curcumin | 4 | 8 | 4 | 8 |
| Synergistic honey + curcumin + pottasium alum | 4 | 8 | 4 | 8 |

**3.2 Time Killing Assay**

The results of the time-killing assay on *P. aeruginosa* after treatment with curcumin, honey, potassium alum, and their combination compared to the base control and imipenem are graphically illustrated in Fig. 1. Both curcumin and honey showed more killing effect on the bacteria in comparison with the imipenem.

In the bacterial killing assay, no bacterial colony was observed at MIC for curcumin, honey, and potassium alum after 16 h, while for the imipenem group, the growth of the bacteria had not yet ceased after this time. Then, the effect of potassium alum, honey, and curcumin on bacterial growth was concluded to be more severe than imipenem according to the statistically significant difference between the treated and control groups ($p < 0.05$). Moreover, the combination of honey, curcumin, and potassium alum showed the most decreasing effect on the bacterial growth so that only in this group, bacteria cease growing after 8 h. The control group showed an obvious increase in bacterial growth. We note that the error bars in Fig. 1 are obtained by the mean fo three repeats for each value ± standard deviation using the *P*-values of $< 0.05$.

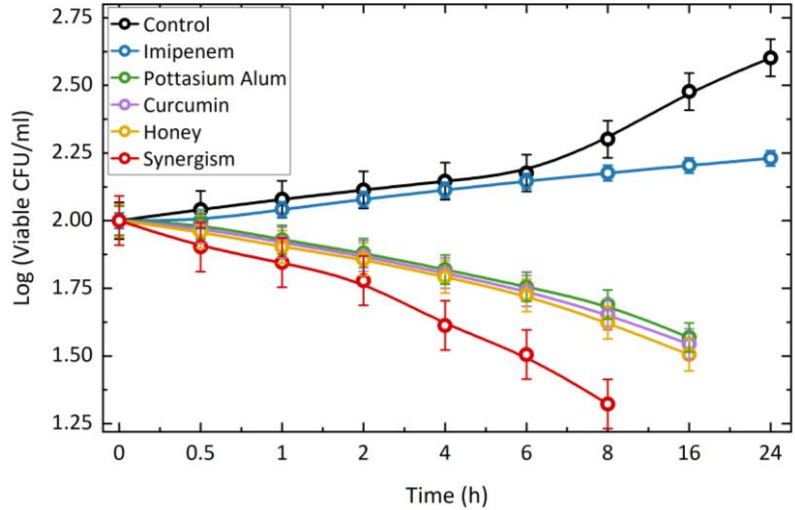

*Figure 1. Time-killing assay for imipenem, curcumin, honey, potassium alum, and the synergy of three active materials (Honey, Curcumin, and potassium alum) were performed against P. aeroginosa. The increasingly synergistic antibacterial effect of honey, curcumin, and potassium alum are clear. Again mention the error bar here*

**3.3 MTT Test**

The MTT test on FSCs at days 1, 3, and 5 post-treatment showed the viability level of these cells in the presence of three concentrations of the drug (synergic curcumin, honey, and potassium alum). The used concentrations are illustrated in Fig. 2(a). Cells' growth in the presence of all ointment concentrations was comparable to the control sample or TCPS. The control sample displayed the viability of untreated FSCs. On the other hand, the MTT test on MEF cells at days 1, 3, and 5 post-treatment showed the viability level of these cells at exposure to three concentrations of the drug. The used concentrations showed no statistically significant relevance to the growth level of both cells (Fig. 2(b), which means, cells' growth in presence of all drug concentrations was comparable to the control sample. The control sample displays the viability of untreated MEF cells. Then, the drug has shown no toxic effect on the two used eucaryotic cell lines. All cells had normal absorption showing that they have normal growth in the presence of all applied concentrations of the prepared ointment.

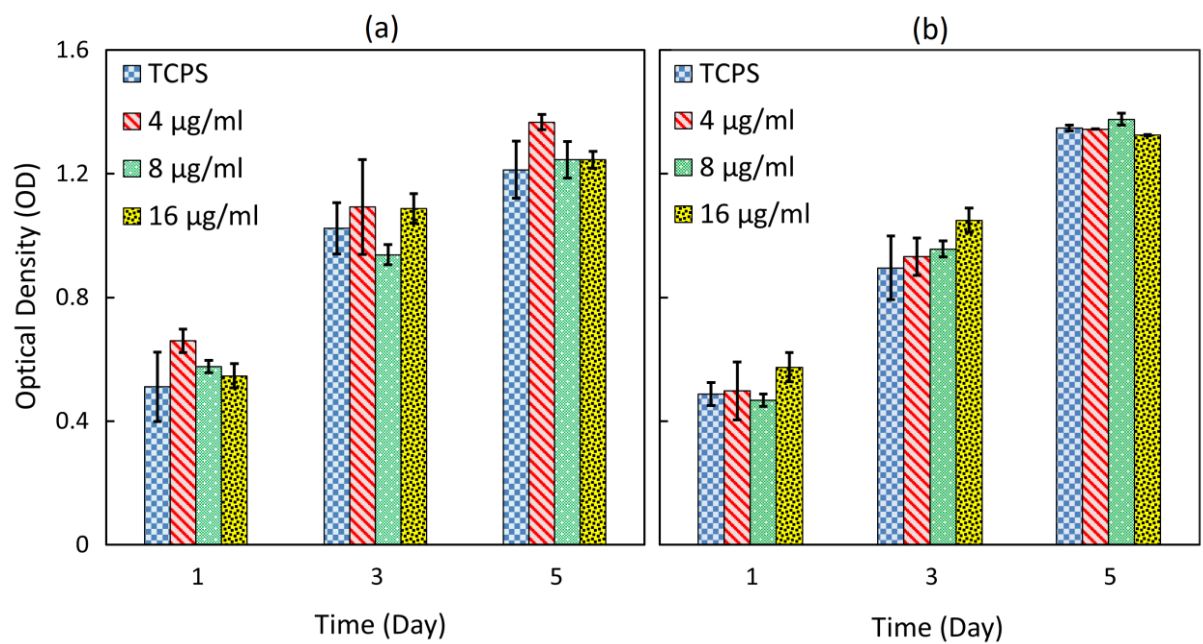

*Figure 2. The cytotoxicity effect of the drug (curcumin, honey, and potassium allum) on FsCs (**a**) and MEFs (**b**) using MTT assay. Cells are cultured for 1, 3, & 5 days with the indicated concentrations (4, 8, & 16 µg/ml) of the drug. The cytotoxicity of the drug is determined based on the percentage of the viable cells in the culture after the scheduled times by measuring the absorption of cell cultures at 540 nm. The data in each column represents three independent experiments ($p < 0.05$).*

### 3.4 Morphological Analysis

For morphological analysis, acridine orange/ethidium bromide (AO/EB) double staining showed the morphology of FSC at days 1, 3, and 5 post-treatment in the presence of three concentrations of the burn ointment (w/v of ointment in culture medium). The used concentrations are illustrated in Fig. 3. Green-stained cells represented viable cells while the apoptotic cells were stained in bright greenish/yellow at early stages and reddish or orange at late apoptosis stages. All concentrations are comparable to the control. The control displayed the morphology of untreated cells (×100 magnification). The staining showed that in treated FSCs, the burn ointment induced no apoptosis after five days of incubation. As shown in Fig. 4, after 48 hours, cellular morphology showed no changes in chromatin condensation or fragmented nuclei. Acridine orange staining showed the normal morphology and growth of MEFs at days 1, 3, and 5 post-treatment in the presence of three concentrations of the burn ointment. The used concentrations are illustrated in Fig. 3. All concentrations are comparable to the control, where it displays the morphology of untreated cells (×100 magnification).

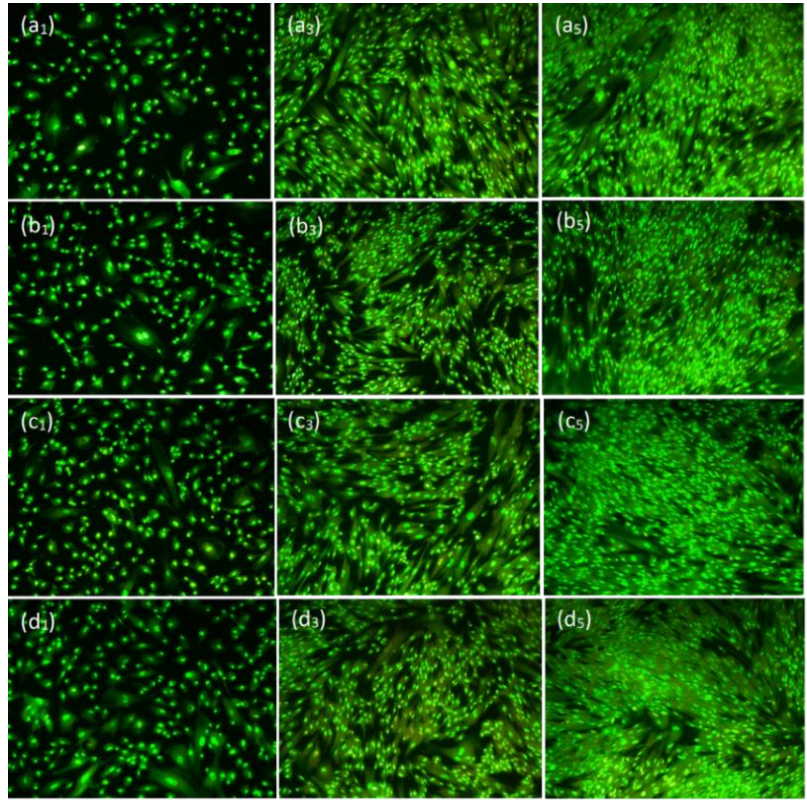

*Figure 3. Akrydince orange staining of FSCs in presence of three concentrations (25 µg/mL, 51 µg/mL, & 102 µg/mL, rows) of the ointment in three times (1, 3, and 5 days, columns) (×100 magnification). AO/EB staining shows an increasing procedure of growth in all groups within five days proving the non-toxicity of the newly-developed burn ointment in different concentrations of 25 (**b**), 51 (**c**), and 102 (**d**) µg/mL. Column (**a**) **is** the control group*

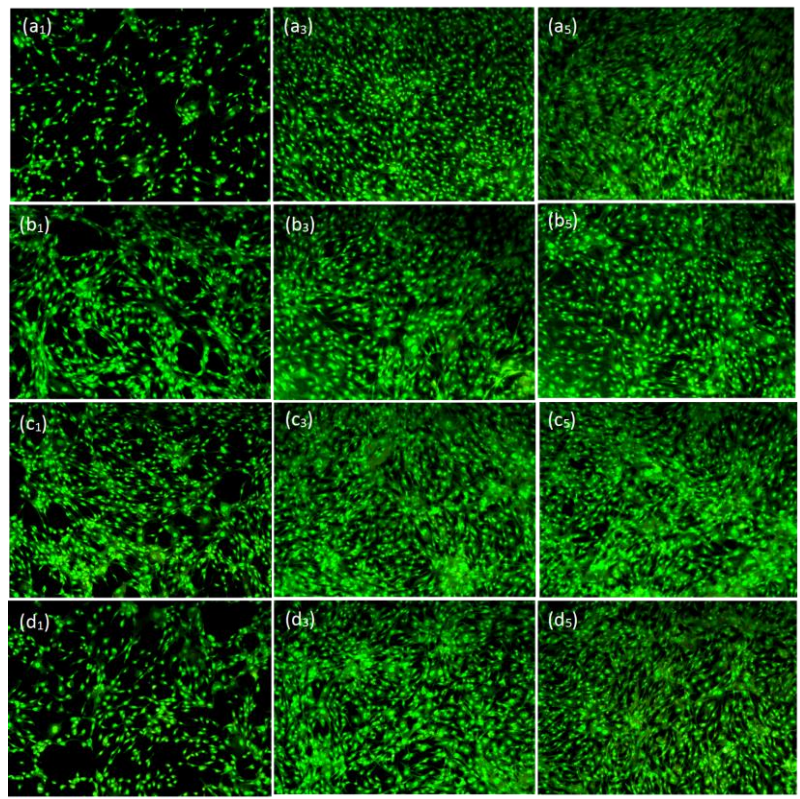

*Figure 4. Akrydince orange staining of MEFs in presence of three concentrations (25 µg/mL, 51 µg/mL & 102 µg/mL, rows) of the ointment in three times (1, 3, and 5 days, columns)(×100 magnification). AO/EB staining shows an increasing*

*procedure of growth in all groups within five days proving the non-toxicity of the newly-developed burn ointment in different concentrations of 25 (**b**), 51 (**c**), and 102 (**d**) µg/mL similar to the control group (**a**).*

### 3.5 *In vivo* Histopathological Evaluation

Histopathological results of the sections stained by H&E (A series on the left side of the figures) and MT (B series on the right side of the figures) staining after 7, 14, 21 days are shown in Figs. 5-7. In the negative control without treatment, the histopathological evaluation at days 7, 14, and 21 post-treatment displayed accumulation of the infiltration of Polymorphonuclear inflammatory cells (PMNs) and granular tissue formation. Where no epidermal layer was formed and a crusty scab had covered all through the wound.

The PMNs infiltration to the unhealthy area of skin was also observed in the positive control group at day 7 post-treatment. On day 14, although a thin layer of epithelial cells was observed, the inflammatory cells were apparent in the unhealthy area of skin and the wound was still covered with a crusty scab. Finally, on day 21 post-treatment, the epithelialization process was completed and the inflammatory cells were significantly decreased in comparison to the negative control group.

The micrographs of the base group (received the base ointment) at days 7, 14, and 21 post-treatment showed similarity to the negative control group displaying a crusty wound with inflammatory and no epiderm layer formation in the wound area. However, the inflammation reaction rate and the number of inflammatory cells at days 14 and 21 post-treatment had significantly decreased. The base group's micrographs showed to be highly similar to the negative control group at 7, 14, and 21 days post-treatment. In this group, the wound area was covered with a crusty scab showing no epidermal formation and high inflammation. However, the amount of the inflammatory responses in the base group considerably decreased at 14 and 21 days post-treatment in comparison to the negative control group.

The histopathological assessments of the treatment group (treated with the testing ointment), revealed a significant decrease in the number of inflammatory cells at day 14 post-treatment in comparison with the other groups. The samples of the treatment group represented the most similarity to the normal skin among all four groups as they had formed a thin epiderm, normal rete ridges, and rejuvenation of skin appendages. Overall, the treatment group had the best healing results among all four groups. Histopathological examination of PC on day 7 proved severe infiltration of inflammatory cells into the defect area. On day 14, inflammation was still evident in the wound area covered by a crusty scab, however, a thin layer of the epithelial cell had formed. Finally, at 21-day post-treatment, the epithelialization process was completed and the inflammatory cells were significantly reduced compared to the negative control group. Generally, histopathological evaluation of the treatment group showed a considerable reduction of inflammation and an increase of epithelization on day 14 compared to other groups. This group revealed more similarity to normal skin having a narrow epidermis, consisting of normal rete ridges, and rejuvenation of skin appendages. Accordingly, the treatment group represented the best epithelization results among all experimental groups. Generally, the highest epithelization and wound healing were detected in the treatment group treated with additives-containing burn wounds and the lowest ones related to the control group (Figs. 5-7 (a-d)).

Since the wound healing procedure is essentially depending on the collagenization process, further investigation was exerted on the effect of the present treatment on wound healing using MT staining. MT staining helped to recognize the synthesis progress of collagen fibers during matrix remodeling and granular tissue (GT) formation.

Accordingly, the intensity of the blue-green color of sections due to MT staining was considered as a function of the relative amount of total deposited collagen molecules and reflecting the advancement of collagenization and tissue remodeling (Figs. 5-7 (e-h)). The results of MT staining indicated that among all, the positive control and treatment groups resulted in the highest collagenization level. On the other hand, the negative control group showed the least amount of collagen fiber formation and deposition in the wound area.

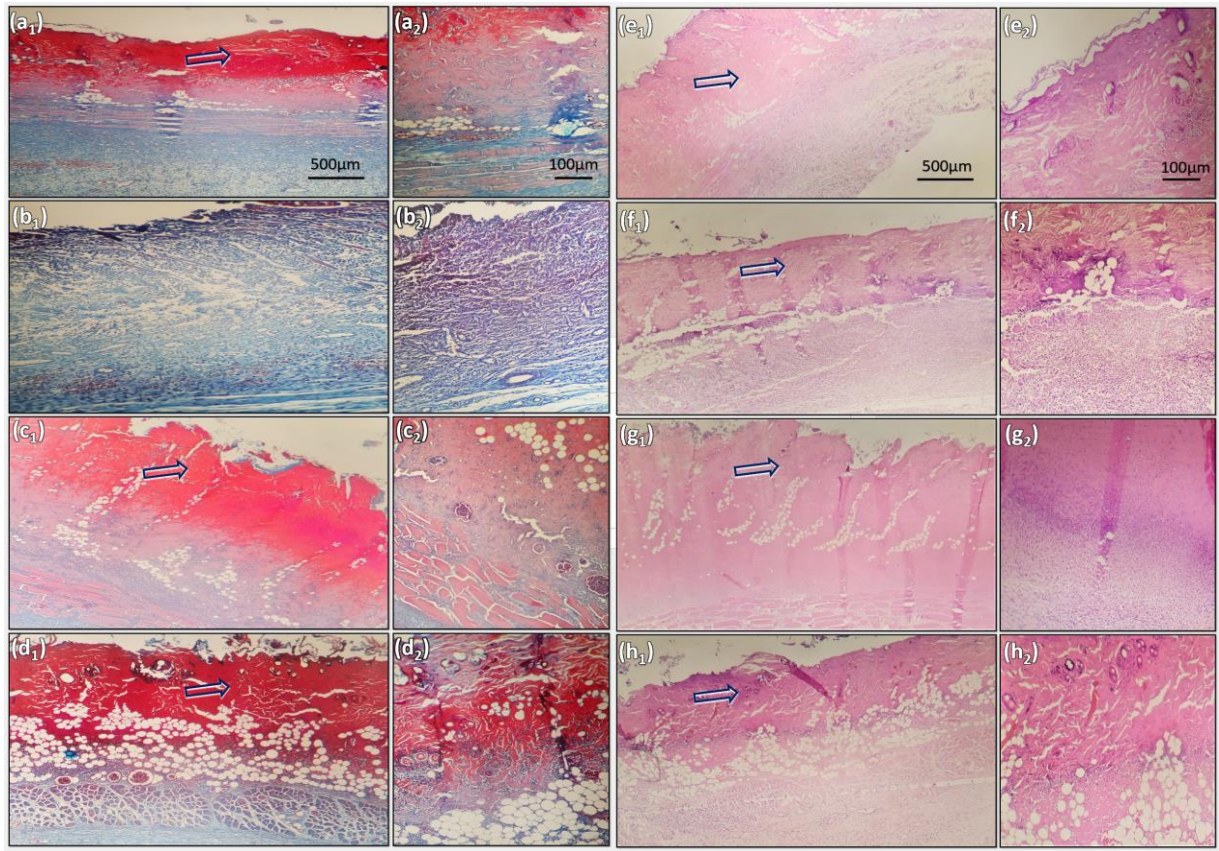

*Figure 5. Microphotographs of healed incision sections at day 7. Hematoxylin and eosin (H&E) stained microscopic sections with X40 (a1-d1) & X100 (a2-d2) magnification in the negative control group: no treatment (a1 & a2), positive control group: treated with zinc oxide ointment (b1 & b2), base group: treated with the base ointment (c1 & c2), and treatment group: burn ointment containing honey and curcumin (d1 & d2). Modified Masson's Trichrome stained microscopic sections with X40 (e1-h1) & X100 (e2-h2) magnification in the negative control group: no treatment (e1 & e2), positive control group: treated with zinc oxide ointment (f1 & f2), base group: treated with the base ointment (g1 & g2), and treatment group: burn ointment containing honey and curcumin (h1 & h2). Arrows: crusty scab.*

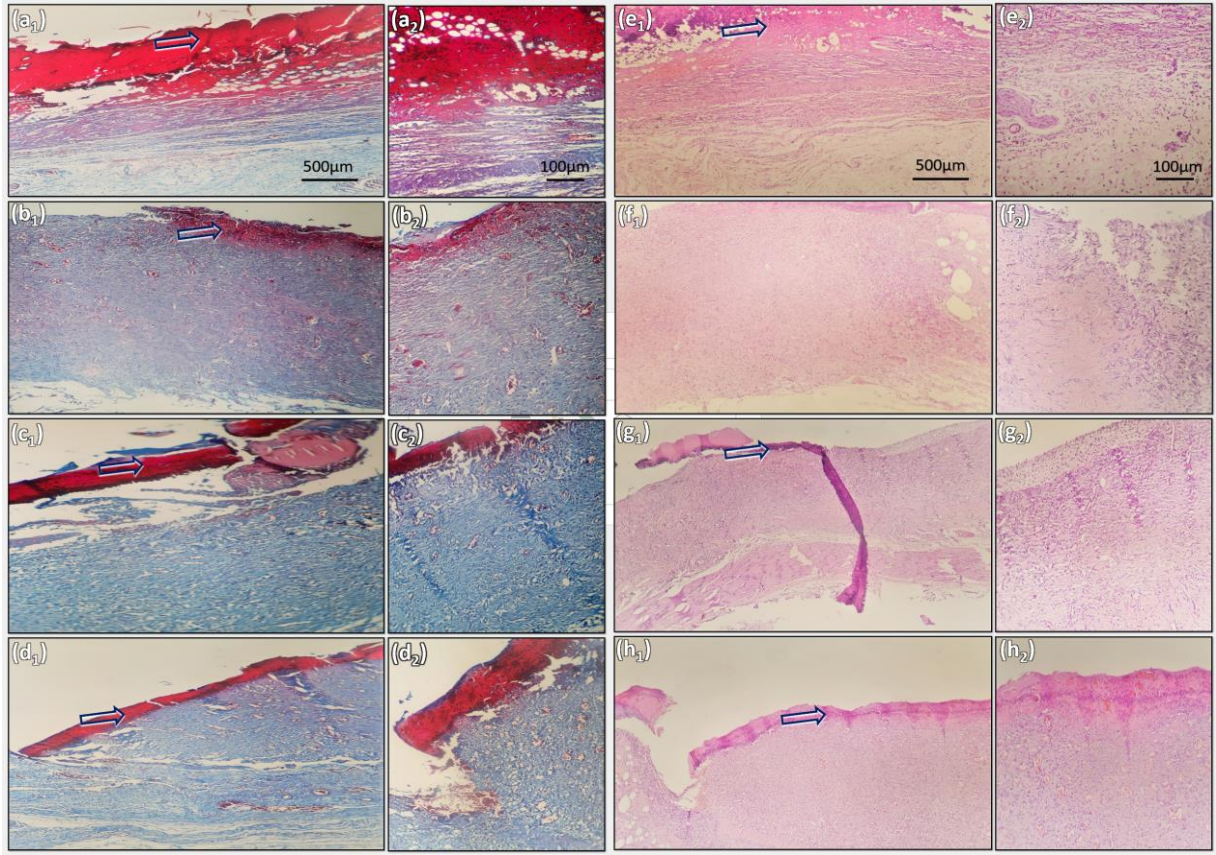

*Figure 6. Microphotographs of healed incision sections at day 14. H&E stained microscopic sections with X40 (**a₁-d₁**) & X100 (**a₂-d₂**) magnification in the negative control group: no treatment (**a₁ & a₂**), positive control group: treated with zinc oxide ointment (**b₁ & b₂**), base group: treated with the base ointment (**c₁ & c₂**), and treatment group: burn ointment containing honey and curcumin (**d₁ & d₂**). Modified Masson's Trichrome stained microscopic sections with X40 (**e₁-h₁**) & X100 (**e₂-h₂**) magnification in the negative control group: no treatment (**e₁ & e₂**), positive control group: treated with zinc oxide ointment (**f₁ & f₂**), base group: treated with the base ointment (**g₁ & g₂**), and treatment group: burn ointment containing honey and curcumin (**h₁ & h₂**). Arrows: crusty scab.*

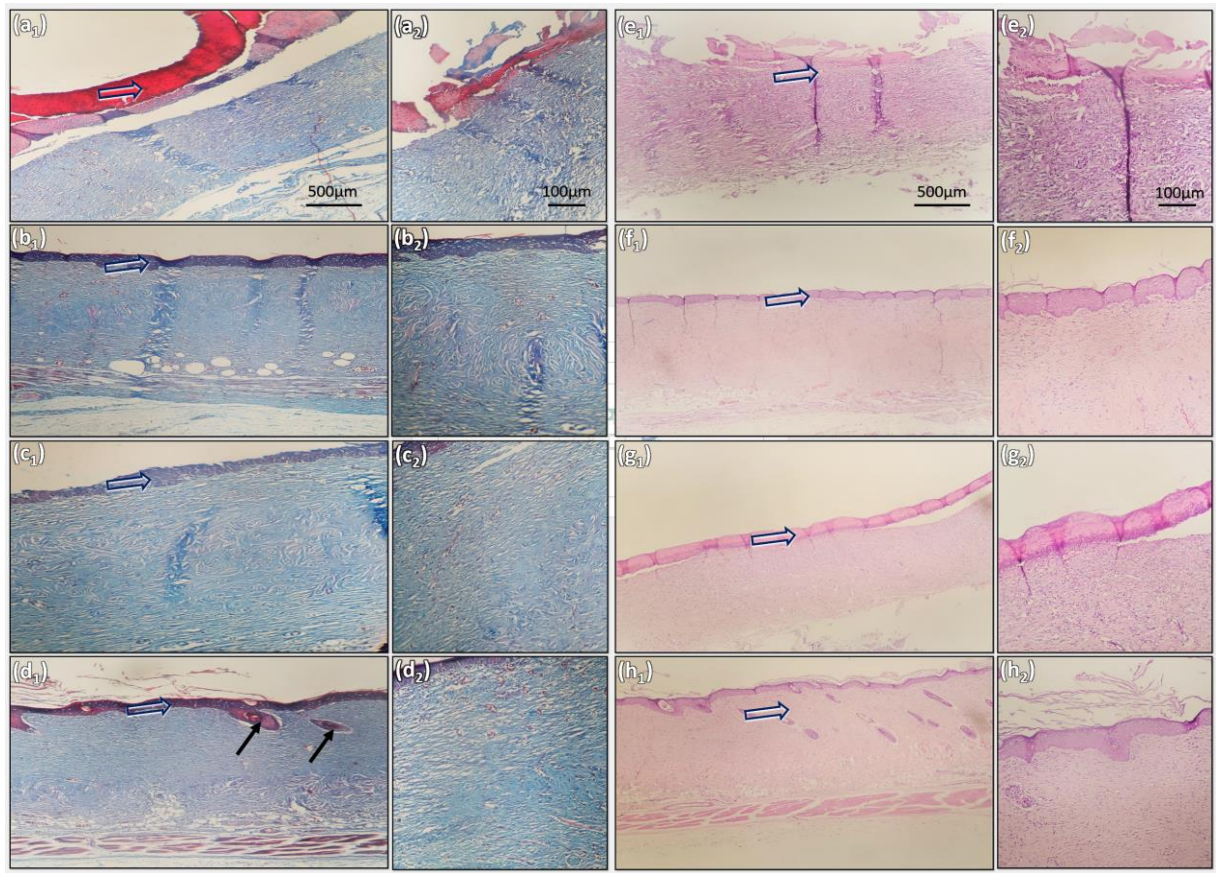

*Figure 7. Microphotographs of healed incision sections at day 21. H&E stained microscopic sections with X40 ($a_1$-$d_1$) & X100 ($a_2$-$d_2$) magnification in the negative control group: no treatment ($a_1$ & $a_2$), positive control group: treated with zinc oxide ointment ($b_1$ & $b_2$), base group: treated with the base ointment ($c_1$ & $c_2$), and treatment group: burn ointment containing honey and curcumin ($d_1$ & $d_2$). Modified Masson's Trichrome stained microscopic sections with X40 ($e_1$-$h_1$) & X100 ($e_2$-$h_2$) magnification in the negative control group: no treatment ($e_1$ & $e_2$), positive control group: treated with zinc oxide ointment ($f_1$ & $f_2$), base group: treated with the base ointment ($g_1$ & $g_2$), and treatment group: burn ointment*

### 3.6 Histomorphometric Assessment

The histomorphometric analysis was conducted 7, 14, and 21 days after creating the skin wound. Re-epithelialization was at the least in the base ointment group and negative control (zinc ointment) group. The wound in these two groups was mostly filled with immature granular tissue ($P < 0.05$). The highest re-epithelialization was observed in the treatment group. Generally, the healing status in the treatment and positive control groups was most similar at day 21. The treatment group showed even much better healing results than the positive control group. The wounds in the test group showed the best appearance and most similar to the natural tissue, the most normal-like thickness of the epidermal layer, the highest level of hair follicle rejuvenation, and enough skin appendages (Table 3).

*Table 3 Histomorphometric analysis of different experimental groups.*

| Group | Collagen density (%) | Angiogenesis/5HPF | Epitheliogenesis score (n = 3) |
|---|---|---|---|
| Negative Control | 14 ± 3 (7) | 8.3 ± 2.5 (7) | 0, 0, 0, 0 (7) |
|  | 20.6 ± 2.5 (14) | 27.6 ± 3.0 (14) | 0, 0, 0, 0 (14) |
|  | 42 ± 3.6 (21) | 34.6 ± 3.5 (21) | 0, 1, 0, 0 (21) |
| Positive Control | 32.3 ± 7.3 ** (7) | 8.6 ± 3 (7) | 0, 0, 0, 0 (7) |
|  | 40 ± 5.2 ** (14) | 32.3 ± 4.5 * (14) | 0, 1, 1, 0 (14) |
|  | 68 ± 2 ** (21) | 15.3 ± 3 * (21) | 3, 3, 4, 3 ** (21) |
| Base | 13.6 ± 4 (7) | 8.3 ± 2 (7) | 0, 0, 0, 0 (7) |
|  | 38.6 ± 6 * (14) | 24.3 ± 2.5 (14) | 0, 0, 0, 0 (14) |
|  | 57.6 ± 5.5 * (21) | 27.3 ± 2.5 * (21) | 2, 1, 1, 1 * (21) |
|  | 18.6 ± 1.5 (7) | 13.3 ± 2.5 (7) | 0, 0, 0, 0 (7) |
| Treatment | 39.6 ± 5.5 ** (14) | 40.3 ± 4.1 * (14) | 1, 0, 1, 0 (14) |
|  | 73.6 ± 5.1 *** (21) | 6.3 ± 1.5 ** (21) | 4, 4, 4, 4 *** (21) |

*, **, ***: These values indicate the significance of the difference between the treatment group versus un-treated group (empty or negative control); * $P \leq 0.05$, ** $P \leq 0.01$, *** $P \leq 0.001$

## 4 Discussion

Most second-degree burn wounds are resulted from hot liquids and importantly affect the social and psychological functions, health, and life quality of patients[39]. On the other hand, burn wounds are highly at risk of infections by *P. aeruginosa* especially in immunocompromised hosts resulting in open and large injuries with necrotic tissues. These opportunistic pathogens carry out threats against public health by transforming into multi-drug resistant (MDR) species which is a major problem especially in underdeveloped and developing countries[12].

Several standard chemical medicines such as silver sulfadiazine have been developed and used for topical treatment of these wounds[1]. But such synthetic products are usually associated with complications and side effects, including neutropenia, erythema, leukopenia, renal toxicity, and delayed wound healing, especially in long-duration consumption. Therefore, various nature-derived therapeutic products are being developed and employed for healing the burn wound and modern scientific methods and clinical trials have come to help to confirm the efficiency of these animal/herbal/microbial-based medications[27]. Several other studies have used potassium alum in wound healing experiments but not many studies have reported its burn healing effect either alone or formulated in a combination[40]. However, the burn wound treatment has remained a not-completely resolved problem. Honey, curcumin, and potassium alum have been individually applied in various platforms (e.g. solutions, gels, ointment, and dressing) and compositions for wound dressing regarding their promising outcome of wound healing, anti-inflammatory, and bactericidal[41]. We took the ointment as it provides the ideal formulation for topical medications due to the least invasive nature and high persistence. Another considerable advantage of curcumin for us was its non-toxicity for healthy cells even in high doses. According to the literature, if a partial-thickness burn wound takes longer than 2 weeks to heal, the risk of leaving scar increases[3]. Then, we hypothesized that honey would be an effective ingredient for its proven acceleration effect on the burn wound healing process.

In several studies, honey has been used with various materials for burn wound healing purposes[1,26,41]. Using honey and milk has been reported to be more effective in burn wound healing than silver. Honey used with garlic has shown synergistic antimicrobial and burn wound healing properties in Wistar rat models by shortened epithelialization and wound contraction time[26]. In another study, different types of Italian honey were examined for their antibacterial properties against *P. aeruginosa* and reported to be 25 – 50 % v/v which is much greater than what we observed here (8 – 16 µg/mL)[41]. Interestingly, honey is suggested to be useful at any stage of healing. However, honey, as well as curcumin, is only recommended for topical (first and second) burn wounds and they are rarely considered for treating severe and deep burn wounds[42]. Accordingly, honey has been compared to mafenide acetate and showed no significantly better pathologic score, concluding to be not effective in deep burn wounds[27]. Also, it has been demonstrated that the re-epithelialization and anti-inflammatory potentials of honey alone have been examined on the third-degree burn wounds infected by *P. aeruginosa* which showed to be not significantly different from those of silver sulfadiazine[13]. Studies address a considerable differentiation in the sensitivity of *P. aeruginosa* isolates to honey. The disparity among these studies may be due to the difference in using honey types and variability of the clinical bacterial isolates[43].

On the other hand, some clinical studies have reported that directly applying the honey to the wound can cause itching or pain as a result of acidity[44]. The ointment platform was selected for our product, and curcumin was used for its analgesic properties to overcome this complication[45]. Also, it has been reported a dose-dependent healing effect for curcumin on the third-degree burn wounds in rat models[46]. Compared with their results, the present findings imply the synergistic healing properties of honey and curcumin[47]. Also, we hypothesized that a burn ointment containing curcumin can prevent biofilm formation which is a crucial factor leading to epithelialization and granulation impairment[48]. Previous studies have proved the antibiofilm properties of curcumin in several bacterial strains including *P. aeruginosa* [15]. Also, Adamczak et al. reported that the preventive properties of curcumin can be strain-specific (working on over 100 strains from 19 species of bacteria and fungi)[49]. Furthermore, they have obtained generally greater sensitivity in Gram-positive bacterias compared to the Gram-negative ones with a very selective antimicrobial activity[49]. Our findings are consistent with the results of Adamczak et al.'s study showing even higher bactericida[50]. In the present study, the MIC of curcumin against clinical isolates and reference strains of *P. aeruginosa* showed to be much less (16 – 32 µg/mL) than their results (62.5 – 5000 µg/mL). Generally, examining different strains of *P. aeruginosa* has shown highly variable sensitivities to curcumin (mostly 30 to >5000 µg/mL). Also, it was shown to be 400 µg/mL in Takrami *et al*.'s study (2019)[51] and 62.5 – 250 µg/mL in the study of Mohammadzamani *et al*. (2020)[14].

Since carbapenems such as imipenem are the selective drugs for the treatment of MDR *P. aeruginosa* isolates, it was used as a comparison reference. The increasingly emerging of carbapenem-resistant *P. aeruginosa* strains as a worldwide concern has motivated us to develop a burn ointment with comparable antibacterial and wound healing properties[52].

By preparing this burn ointment containing curcumin, honey, and potassium alum, we have successfully developed a therapeutic product for the topical treatment of second-degree burn wounds that not only accelerate the healing process of the thermal injury but also prevent the infection. Our product shows no apoptotic effect on the experimented eucaryotic cells in vitro confirmed by both quantitative (MTT) and qualitative (AO staining) examinations. Also, the histological analyses confirmed the wound healing properties of the newly-prepared burn ointment. Then, we hopefully expect that our ointment exerts minimized side-effects including pain experience and allergy in clinical examinations. In a similar study by Abbas *et al*., the synergistic bactericidal and healing properties of curcumin have been shown in association with chitosan in rabbit models[11]. According to their histopathological examinations, the control groups showed delayed healing while the group treated with the prepared burn ointment containing curcumin and honey showed a completely formed epidermis layer. However, potassium alum used in our study has increased the efficacy of the ointment.

Despite all advantages that such tradition-originated therapeutic products have shown for treating burn lesions, their probable sensitivity to the geographic season and location, and exhibition of batch-to-batch variation are challenges to be resolved for developing standard controlled productions[3]. Some mechanisms are counted to be underlying the antibacterial properties of curcumin, honey, and potassium alum that are finely summarized by Adamczak *et al*[15]. As an instance, Mohammadzamani *et al*. have reported that honey in combination with cinnamaldehyde and carvacrol reduces the expressions of exoS and ampC genes in *P. aeruginosa*.[14] However, further studies are still required to describe the underlying detailed mechanisms. On the other hand, the wound type and location are involved in the efficacy of wound healing activities of honey. Then, larger and longer studies and well-designed randomized trials are recommended for determining the statistical significance of curcumin, honey, and potassium alum efficacy in curing different burn wounds.

## 5 Conclusion

This study aimed to assess the healing properties of a newly-prepared burn ointment containing curcumin, honey, and potassium alum on second-degree burn wounds in rats. Our results showed substantial antibacterial and wound healing properties with no cytotoxicity. The significant efficiency of this burn ointment in healing the experimentally-created second-degree wound besides its potential of relieving pain and inflammation suggests this product as a promising therapeutic and palliative care for first and secondhand burn wounds. Although, it is recommended to be further clinically studied.

**Acknowledgments** The authors thank Terita Sana Darman Co. and Mr. Younes Mollarahman for their moral support from this research.

**Declarations**

**Funding** The funding for the present study was provided by Terita Sana Darman company.

**Conflict of interest** The authors declare they have no interests that might be interpreted as influencing the article.

**Availability of data and material** Raw data are available on request from the authors.

**Consent to participate** Not applicable.

**Code availability** Not applicable.

**Author contributions** All authors contributed to the study conception and design. All authors read and approved the final manuscript.

in Pseudomonas aeruginosa burn isolates in Tehran. *Jundishapur J. Microbiol.* **6**, 162–165 (2013).